\documentclass[
  aps,
  prd,
  reprint,          
  twoside,          
  printing,
  superscriptaddress,
  nofootinbib,
  longbibliography
]{revtex4-2}

\usepackage[T1]{fontenc}
\usepackage{lmodern}
\usepackage{amsmath,amssymb,mathtools, bm}
\usepackage{graphicx}
\usepackage{dblfloatfix}

\usepackage{booktabs}
\usepackage{multirow}
\usepackage{siunitx}
\usepackage{xcolor}
\usepackage{ragged2e}
\usepackage{caption}
\usepackage{hyperref}
\usepackage{cleveref}
\usepackage{microtype}
\usepackage{enumitem}
\usepackage{comment}

\usepackage{subcaption}
\usepackage{placeins}
\definecolor{darkgold}{HTML}{9A7000}

\hypersetup{
  colorlinks=true,
  linkcolor=darkgold,
  citecolor=darkgold,
  urlcolor=darkgold
}
\newcommand{\dd}{\mathrm{d}}

\begin{document}

\title{Direct Detection of Millicharged Particles from Evaporating Primordial Black Holes}

\author{Biprajit Mondal}
\email{biprajitm@iisc.ac.in}
\affiliation{Centre for High Energy Physics, Indian Institute of Science, C. V. Raman Avenue, Bengaluru 560012, India}

\date{\today}

\begin{abstract}
We investigate the direct detection of millicharge particles produced via Hawking evaporation of neutral primordial black holes~(PBHs). As they travel through the Galactic magnetic field and then Earth's overburden, they lose energy before reaching an underground detector. Using electron recoil signals, we obtain constraints on the mass and charge parameter space for XENONnT, JUNO and Super-K for evolved PBH populations with initial mass $ M_i=10^{15}~\rm g$ and $ M_i=8\times10^{15}~\rm g$ with initial abundances $f_i^{\rm PBH}=3.9\times10^{-7}$ and $f_i^{\rm PBH}=3.7\times10^{-4}$, respectively. In the relevant parameter space, this model can provide strong constraints in the high mass region.
\end{abstract}

\maketitle

\section{Introduction}
\label{sec:intro}

A cosmological relic like a primordial black hole~(PBH) is a source of the Standard Model~(SM) particles as well as Beyond the Standard Model~(BSM) particles. If PBHs~\cite{Bertone:2004pz,10.1093/mnras/152.1.75,Green:2020jor,Carr:2020gox,Cirelli:2024ssz,Saha:2024ies,Saha:2025wgg,Tamta:2024pow,Saha:2021pqf,Laha:2019ssq,Dasgupta:2019cae,Laha:2020ivk,Laha:2020vhg,Ray:2021mxu} and  millicharge particles~(MCPs)~\cite{PhysRevD.41.2388, McDermott:2010pa,Chuzhoy:2008zy,Prinz_1998,DelNobile:2015bqo,Liu:2019knx,Pospelov:2020ktu, Laha:2018zav} exist, MCPs can be emitted through Hawking evaporation of PBHs. MCPs produced by different sources, such as supernovae and cosmic ray interactions, have been searched for in direct detection through their interactions with electrons and nuclei. Hawking evaporation of PBHs with mass $\rm M_{\rm PBH}=\mathcal{O \rm (10^{15}g)}$ provides a substantial flux of MCPs with masses in the sub-MeV to MeV range, which can be detected in direct detection experiments.

 In this work, we investigate whether MCPs emitted by PBHs can reach the XENONnT~\cite{XENON:2026rbz}, JUNO~\cite{JUNO:2020hqc} and Super-K~\cite{Du:2022hms} detectors and produce a detectable signal.  XENONnT is useful because its low recoil threshold enhances the sensitivity to the large scattering cross section at low recoil energies, with even smaller recoil energies than nuclei, and can therefore produce observable electron recoil signals, whereas JUNO and Super-K probe higher energy recoil events and benefit from their large exposures. We generalise the standard PBH evaporation framework from a neutral particle~\cite{Calabrese:2021src} to a particle coupled to the photon through a small electric charge.

Using the MCP spectrum and its galactic transport, we calculate the MCP flux at the detector and derive exclusion limits for XENONnT and Super-K, along with the projected sensitivity for JUNO. We compare our results with existing constraints~\cite{Magill:2018tbb,Chang:2018rso,Li:2024pcp,Vinyoles:2015khy}.

The paper is organised as follows. In \cref{sec:emission}, we describe the production of millicharge particles through PBH evaporation. In \cref{sec:flux}, we derive the Galactic flux, discuss its propagation to Earth and the detector. In \cref{sec:result}, we present the results and conclude here.

\section{Hawking emission}
\label{sec:emission}
We consider two kinds of electrically neutral and nonrotating PBHs with initial masses $\rm 10^{15}~\rm g$, and $8\times10^{15}~\rm g$, with the initial abundances  $f_i^{\rm PBH}$ to be $3.9\times10^{-7}$ and $3.7\times10^{-4}$, respectively~\cite{Carr:2020gox,koechler2026constrainingdarkmatterusing}. The PBHs emit MCPs $\chi$, with electric charge
$q_\chi=\epsilon e$, together with its antiparticle $\bar{\chi}$. The Hawking temperature of a PBH is~\cite{Calabrese:2021src}
\begin{equation}
k_B T_{\rm H}
\simeq
10.6\,\mathrm{MeV}
\left(
\frac{10^{15}\,\mathrm{g}}{M_{\rm PBH}}
\right).
\label{eq:hawking_temperature}
\end{equation}

The differential emission spectrum per unit time per unit energy for a spin-$1/2$ particle is
\begin{equation}
\frac{\dd^2 N^{\rm emission}_\chi}{\dd T_\chi\,\dd t}
=
\frac{g_\chi}{2\pi\hbar}
\frac{
\Gamma(T_\chi,m_\chi,M_{\rm PBH}(t_0;m_\chi))
}{
\exp\!\left[
\frac{T_\chi+m_\chi}
{k_B T_{\rm H}(M_{\rm PBH}(t_0;m_\chi))}
\right]+1
}.
\label{eq:hawking_spectrum}
\end{equation}
where $T_\chi$ is the kinetic energy of the emitted particle, $g_\chi=4$ is the total number of internal degrees of freedom of the Dirac field, $k_B$ is the Boltzmann constant, $\hbar$ is the reduced Planck constant, and $\Gamma(T_\chi,M_{\rm PBH}(t_0))$ is the corresponding greybody factor. The differential emission spectrum has unit of $\rm MeV^{-1}s^{-1}$. We compute the emission spectrum using \texttt{BlackHawk}~\cite{Arbey:2019mbc}. Once the PBHs have evolved to their present-day~($t_0$) masses, where $t_0=13.8~{\rm Gyr}$ is the age of the universe, they are treated as a source of the instantaneous emission spectrum.  We, therefore, neglect the explicit time dependence of the source, $\dd^2N^{\rm emission}_\chi/(\dd T_\chi,\dd t)[M_{\rm PBH}(t)]$, because the Galactic propagation timescale, which is of order $\rm Myr$, is much shorter than the PBH mass evolution timescale ($\sim\rm Gyr$). While evolving the PBHs using \texttt{BlackHawk}, we also included the MCP emission channel together with the SM channel. The evaporation rate, therefore, depends on $m_\chi$. For $m_\chi \lesssim k_B T_{\rm H}$, MCP emission is efficient, whereas for $m_\chi \gtrsim k_BT_{\rm H}$, MCP emission becomes increasingly suppressed. Consequently, for the  $\rm 10^{15}\mathrm{g}$ PBH case, which is much hotter than the other one, MCP emission is more efficient and has a greater impact on its mass evolution.

\section{Millicharge flux from PBH evaporation}
\label{sec:flux}

We divide the PBHs population into two regions: the confinement zone and the outer halo. The confinement zone is modelled as a cylinder with $|z|<3~{\rm kpc}$~\cite{Dunsky:2018mqs} for the vertical extent, and $R<20~{\rm kpc}$ for the radial extent. The charge particles after being emitted by the PBHs situated in the confinement region contribute to the inner flux. The number of PBHs inside the confinement region is given by,
\begin{equation}
N_{\rm PBH}^{\rm in}
=
\frac{f_{\rm PBH}}{M_{\rm PBH}}
\int_{V_{\rm conf}}
\rho_{\rm NFW}(|\mathbf{x}|)\mathrm{d}^3x ,
\end{equation}
where $V_{\rm conf}$ is the volume of the confinement region and $\mathbf{x}=(R,\phi,z)$ is the galactic position in cylindrical coordinates.

PBHs outside the confinement zone, $R\geq20~{\rm kpc}$ or $|z|\geq3~{\rm kpc}$ contribute to the outer flux. The total local interstellar~(LIS) flux is then written as,
\begin{equation}
\frac{d\Phi_{\chi}^{\rm LIS}}{dT_{\chi}}
=
\frac{d\Phi_{\chi}^{\rm in}}{dT_{\chi}}
+
\frac{d\Phi_{\chi}^{\rm out}}{dT_{\chi}}.
\label{eq:flux_equation}
\end{equation}

The galactic halo extends up to $200~{\rm kpc}$. For the present analysis, we truncate the outer source population at $50~{\rm kpc}$, since sources beyond this distance give a negligible contribution to the flux in our setup.\\[1pt]

\subsection{Inner flux}
\label{subsec:inner_flux}

We adopt the leaky-box model~\cite{Harnik:2020ugb} for the confinement zone defined above and apply the galactic transport equation for the propagation of the steady-state MCPs in this region, given by~\cite{Harnik:2020ugb,Kachelriess:2020ams,Strong_2007}

\begin{equation}
\begin{split}
&
\frac{\partial}{\partial p_\chi}
\left[
\left(
b_p
+
\frac{p_\chi}{3}
\nabla\cdot\mathbf{V}_{\rm adv}
\right)
N_\chi
+
p_\chi^2 D_{pp}(p_\chi,W)
\frac{\partial}{\partial p_\chi}
\left(
\frac{N_\chi}{p_\chi^2}
\right)
\right]
\\
&
+
\nabla\cdot
\left[
\frac{1}{2}{D(T_\chi,W)}
\nabla N_\chi
\right]
-
\nabla\cdot
\left[
\mathbf{V}_{\rm adv}\,N_\chi
\right]
+
Q_\chi(p_\chi,\mathbf{x})
=0
\label{eq:full_transport}
\end{split}
\end{equation}
where $N_\chi(p_\chi,\mathbf{x})=\frac{dn_\chi}{dp_\chi}$ has units ${\rm cm^{-3}\,MeV^{-1}}$. The first term is the momentum space evolution of the MCPs in the magnetic field. The quantity $b_p$ defines momentum loss rates like ionisation, Coulomb, and radiative loss when MCPs interact with matter, and the term $\frac{p_\chi}{3}\nabla\cdot\mathbf{V}_{\rm adv}$ accounts for the adiabatic momentum loss associated with the galactic wind. For the synchrotron loss, the effective galactic magnetic field strength is $B=1~\mu{\rm G}$. The radiation energy density for inverse Compton losses is taken to be $U_{\rm rad}=1~{\rm eV\,cm^{-3}}$~\cite{Krumholz:2022oja}. We model the galactic wind as a vertical outflow away from the galactic disk. Thus, $\nabla\cdot \mathbf{V}_{\rm adv}=V_0/H$, and $V_0=30 \rm~ km/s$ is the advection speed. The coefficient $D_{pp}=\frac{2p_\chi^2v_A^2}{9D(T_\chi,W)}$  describes stochastic reacceleration due to interactions with Alfvén-wave turbulence, where $v_A=30~\rm km/s$ is the Alfvén speed. We consider the magnetic field to be $B_0=1~\mu{\rm G}$~\cite{Kachelriess:2020ams}, which determines the Larmor radius and the corresponding resonant wave number for wave resonance calculation. The second term is the spatial diffusion, with $D(T_\chi,W)$ denoting the spatial diffusion coefficient. If we define,

\begin{equation}
D(T_\chi)=D_0\,\beta(T_\chi)
\left(
\frac{\mathcal{R}(T_\chi)}{\mathcal{R}_0}
\right)^{1/3},
\label{eq:diffusion_coefficient}
\end{equation}
then $D(T_\chi,W)=D(T_\chi)
\frac{W_{\rm K}(k)}{W(k)}$,
where $W(k)$ denotes the turbulent wave spectrum and $W_{\rm K}(k)$ is the corresponding unperturbed Kolmogorov spectrum. The third term is the transport of MCPs by the same vertical galactic wind. We take $D_0=3\times10^{28}~{\rm cm^2\,s^{-1}}$, and
$\mathcal{R}_0=1~{\rm GV}$. The rigidity of an MCP
with electric charge $q_\chi=\epsilon e$ is,
\begin{equation}
\mathcal{R}(T_\chi)
=
\frac{p_\chi(T_\chi)}{\epsilon e},
\qquad
p_\chi(T_\chi)
=
\sqrt{T_\chi\left(T_\chi+2m_\chi\right)},
\label{eq:rigidity}
\end{equation}
while its velocity is,
\begin{equation}
\beta_\chi(T_\chi)
=
\frac{p_\chi(T_\chi)}{T_\chi+m_\chi}.
\label{eq:beta}
\end{equation}

For PBHs located within the confinement region, the source term
$Q^{\rm in}_{\chi}(p_\chi,\mathbf{x})$ is determined by their spatial distribution and
Hawking emission spectrum,
\begin{equation}
Q_\chi^{\rm in}(p_\chi,\mathbf{x})
=
n_{\rm PBH}(|\mathbf{x}|)
\frac{d^2N^{\rm emission}_\chi}{dp_\chi\,dt}
\label{eq:source_term}.
\end{equation}
Here, $n_{\rm PBH}(\mathbf{x}) =\frac{f_{\rm PBH}}{M_{\rm PBH}}\rho_{\rm NFW}|\mathbf{x}|$ is the PBH number density at position $\mathbf{x}$ with units of $\mathrm{cm^{-3}}$. Together with Eq.~\eqref{eq:full_transport} and the particle escape boundary condition, $\frac{v_\chi}{4}N_\chi^{\rm in}=-\frac{D(T_\chi,W)}{2}\partial_\perp N_\chi^{\rm in}
$, where $v_\chi(T_\chi)=c\,\beta_\chi(T_\chi)$, we solve for $N_\chi^{\rm in}$ and the corresponding differential flux at the Earth is,
\begin{equation}
\frac{d\Phi_\chi^{\rm in}}{dT_\chi}
=
c
N_\chi^{\rm in}
\left(
p_\chi,
\mathbf{x}_\oplus
\right).
\label{eq}
\end{equation}
where $\mathbf{x}_\oplus$ denotes the position of the Earth.\\[1pt]

\subsection{Outer flux}
\label{subsec:outer_flux}

For the outer flux, the situation changes. Outside the confinement region, the magnetic field configuration is different, and we cannot apply the same galactic transport equation there. Once the particles enter the confinement region, their propagation is described by Eq.~\eqref{eq:full_transport}. If the emitted particles had sufficient rigidity, as in the case of UHECRs, their propagation could be approximately ballistic, and they could penetrate the confinement region with relatively small magnetic deflections. In our case, however, the MCPs have much lower rigidities; their propagation is significantly affected by the magnetic field outside the confinement region. The PBHs outside the confinement region are assumed to follow the same NFW distribution. The injection of the particles through the magnetic field is followed numerically using {\textsc CRPropa}~\cite{AlvesBatista:2022, Merten:2017}. We adopt an isotropic diffusive propagation module \texttt{DiffusionSDE} in \textsc{CRPropa}. For the magnetic field configuration, we supplement the regular galactic JF12 field with a circumgalactic magnetic field component normalised to $B_{\rm rms}^{\rm CGM}=0.1~\mu{\rm G}$~\cite{Pakmor:2020}. This is useful for our transport framework, where the effective scattering of MCPs is described by a rigidity dependent diffusion coefficient $D(T_\chi)$, without the need to track numerically expensive individual gyro-orbits across galactic distances~\cite{Perri:2025tvq}.

To propagate MCPs from the outer halo into the confinement region, we sample their initial positions according to the NFW distribution outside the confinement cylinder and within a galactocentric radius of $50~{\rm kpc}$. A particle is accepted when it first enters the confinement region, $R<20~{\rm kpc}$ and $|z|<3~{\rm kpc}$, and its entry position and propagation time are recorded. Particles escaping beyond $50~{\rm kpc}$ are discarded. The particle propagation timescale ranges from Myr to $2~\rm Gyr$. Once an MCP enters the confinement region successfully, its propagation toward Earth is described by the galactic transport equation~\eqref{eq:full_transport}. The current due to the MCPs at the boundary of the confinement region is given by,
\begin{equation}
J_{\rm in}(p_\chi,\mathbf{x}_b)
=
\frac{N^{\rm out}_{\rm PBH}}{A_b}\frac{N_{\rm accepted}(b;p_\chi,\epsilon)}{N_{\rm simulated}}
\frac{d^2N_\chi^{\rm emission}}{dp_\chi\,dt}
\label{eq:outer_boundary_source}
\end{equation}
where $\mathbf{x}_b$ denotes a position on the boundary of the confinement region, $A_b$ is the area of the corresponding boundary element. $N_{\rm accepted}$ is the number of simulated MCP trajectories entering boundary element $b$ and $N_{\rm simulated}$ is the total number of injected trajectories.

There is still a probability that MCPs escape from the confinement region due to its leaky-box behaviour. The transport equation~\eqref{eq:full_transport} inside the confinement region, now with $Q_\chi=0$, is then
solved with the boundary condition~\cite{Fick:1855},
\begin{equation}
\begin{aligned}
\frac{v_\chi(T_\chi)}{4}\,
N_\chi^{\rm out}(p_\chi,\mathbf{x}_b)
&- J_{\rm in}(p_\chi,\mathbf{x}_b)
\\
&= -\frac{D(T_\chi,W)}{2}\,
\partial_{\perp}N_\chi^{\rm out}(p_\chi,\mathbf{x}_b).
\end{aligned}
\label{eq:outer}
\end{equation}

\begin{figure}[t]
\centering
\includegraphics[width=0.92\linewidth]{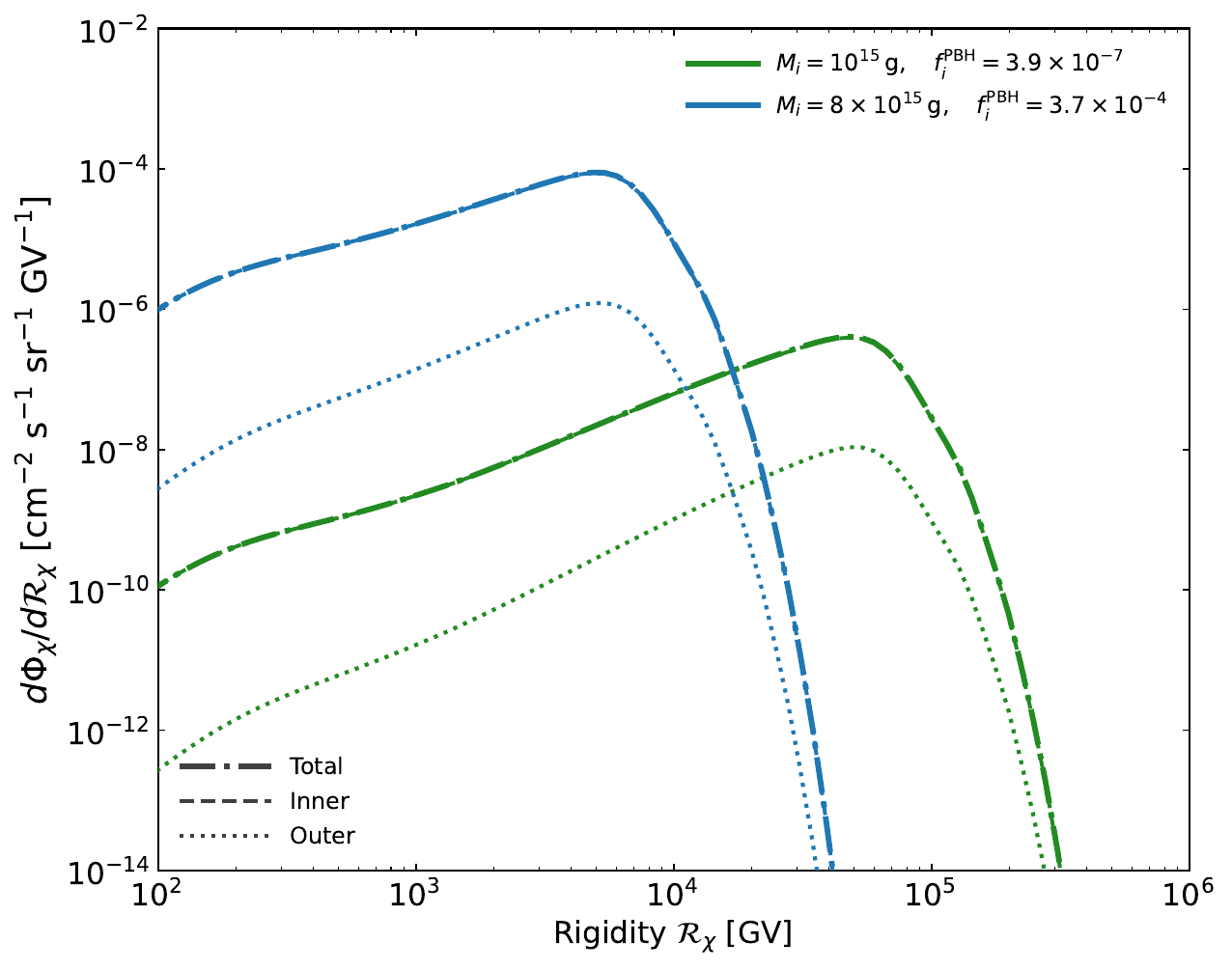}
\caption{\justifying
Local interstellar MCP flux spectra for the benchmark parameters
$m_\chi=1~{\rm MeV}$ and $\epsilon=10^{-6}$. Here, $\rm M_i$ denotes the initial PBH mass with the corresponding initial abundance $f_i^{\rm PBH}$.
The inner-PBH contribution dominates the total flux, accounting for
approximately $97\%$ of the total flux for this benchmark.
}
\label{fig:mcp_flux_inner_outer}
\end{figure}

\subsection{Solar Modulation}
\label{subsec:solar_mod}

The magnetic field of the solar wind can affect the MCP flux before they reach Earth. Particles with higher rigidity are less affected by solar modulation. The net differential flux that can reach Earth is~\cite{Dunsky:2018mqs},
\begin{equation}
\frac{d\Phi_{\chi}^{\rm TOA}}{dT_\chi}
=
\exp\left[
-\frac{\eta(t)}
{\beta_\chi\,g(R_\chi)}
\right]
\frac{d\Phi_{\chi}^{\rm LIS}}{dT_\chi},
\label{eq:solar_module}
\end{equation}

where $\eta(t)=0.3~\mathrm{GeV}$ and
$$
g(\mathcal{R})=
\begin{cases}
\mathcal{R}, & \mathcal{R}>\mathcal{R}_c,\\
\mathcal{R}_c, & \mathcal{R}\leq\mathcal{R}_c,
\end{cases}
$$

with $\mathcal{R}_c=1~\mathrm{GV}$.

\subsection{Overburden}
\label{subsec:overburd}

Before reaching the underground detector, the particles pass through the overburden, where they lose energy through electromagnetic interaction. The initial momentum of a particle hitting the overburden is,

\begin{equation}
u_i
=
\sqrt{\left(1+\frac{T_{\chi,i}}{m_\chi}\right)^2-1}.
\end{equation}

The particle's final momentum $u_f$, after
propagating through the overburden, is evaluated from
\begin{equation}
X(\theta)
=
\frac{m_\chi}{\epsilon^2}
\int_{u_f}^{u_i}
\frac{\beta_\chi du}{S(u,m_\chi)} .
\label{eq:angular_overburden}
\end{equation}
Here, $S(u,m_\chi)$ denotes the mass stopping power.

The corresponding column depth is,

\begin{equation}
X(\theta)
=
X_{\rm vertical}\frac{D(\theta)}{d},
\end{equation}

where

\begin{equation}
D(\theta)
=
\sqrt{(R_\oplus-d)^2\cos^2\theta+d(2R_\oplus-d)}
-
(R_\oplus-d)\cos\theta .
\end{equation}

Here, $R_\oplus$ is the Earth radius, $d$ is the detector's physical vertical depth, and $X_{\rm vertical}$ is the vertical column depth measured in $\rm m.w.e.$ 

For JUNO and Super-K, we use the finite mass Bethe-Bloch equation given by,

\begin{equation}
\begin{aligned}
S_{\rm BB}(u,m_\chi)
&=
K\frac{Z}{A}\frac{1}{\beta_\chi^2(u)}
\biggl[
\frac{1}{2}
\ln\left(
\frac{2m_e u^2 E_{\rm max}(u,m_\chi)}
{I^2}
\right)
\\
&\hspace{3.5cm}
-\beta_\chi^2(u)
-\frac{\delta(u)}{2}
\biggr].
\end{aligned}
\label{eq:BB}
\end{equation}

Here, $Z/A=0.5$, $I=136.4~{\rm eV}$ is the mean excitation energy of standard rock. $\delta(u)$ is the density effect correction~\cite{ParticleDataGroup:2026mpi}.

Due to the low recoil threshold of XENONnT, some MCPs with $\beta_\chi<0.1$ can also contribute to the signal. In this velocity region, the Bethe-Bloch expression becomes less reliable. For this reason, we use the NIST PSTAR electronic stopping-power data~\cite{Berger:PSTAR} for $\mathrm{SiO_2}$ as an approximation of the standard rock. We apply the following conditions: (i) if an MCP enters the overburden with $\beta_\chi<0.1$, we use tabulated PSTAR stopping power; (ii) if it initially has $\beta_\chi\geq0.1$, we use the Bethe-Bloch expression and switch to PSTAR if its velocity falls below $\beta_\chi=0.1$ during propagation, and (iii) if it maintains $\beta_\chi\geq0.1$ throughout its propagation, only the Bethe-Bloch stopping is applied.

For the PSTAR case, we approximate the MCP mass stopping power by,
\[
-\frac{dT_\chi}{dX}
\simeq \epsilon^2 S_{p}^{\rm PSTAR}(u),
\]
where $S_{p}^{\rm PSTAR}$ is the proton electronic stopping power with $T_p=m_p(\gamma\beta_\chi-1)$.

For a particle that reaches the detector with kinetic energy $T_{\chi,f}$, we calculate the maximum electron recoil energy it can produce,

\begin{equation}
E_e^{\max}
=
\frac{
2m_e\left(E_{\chi,f}^2-m_\chi^2\right)
}{
m_\chi^2+m_e^2+2m_eE_{\chi,f}
}.
\end{equation}

If \(E_e^{\max}\geq E_e^{\rm thr}\), the particle is kinematically capable of producing a detectable recoil and contributes to the signal rate.
%




\begin{figure}[t]
\centering
\includegraphics[width=0.92\linewidth]{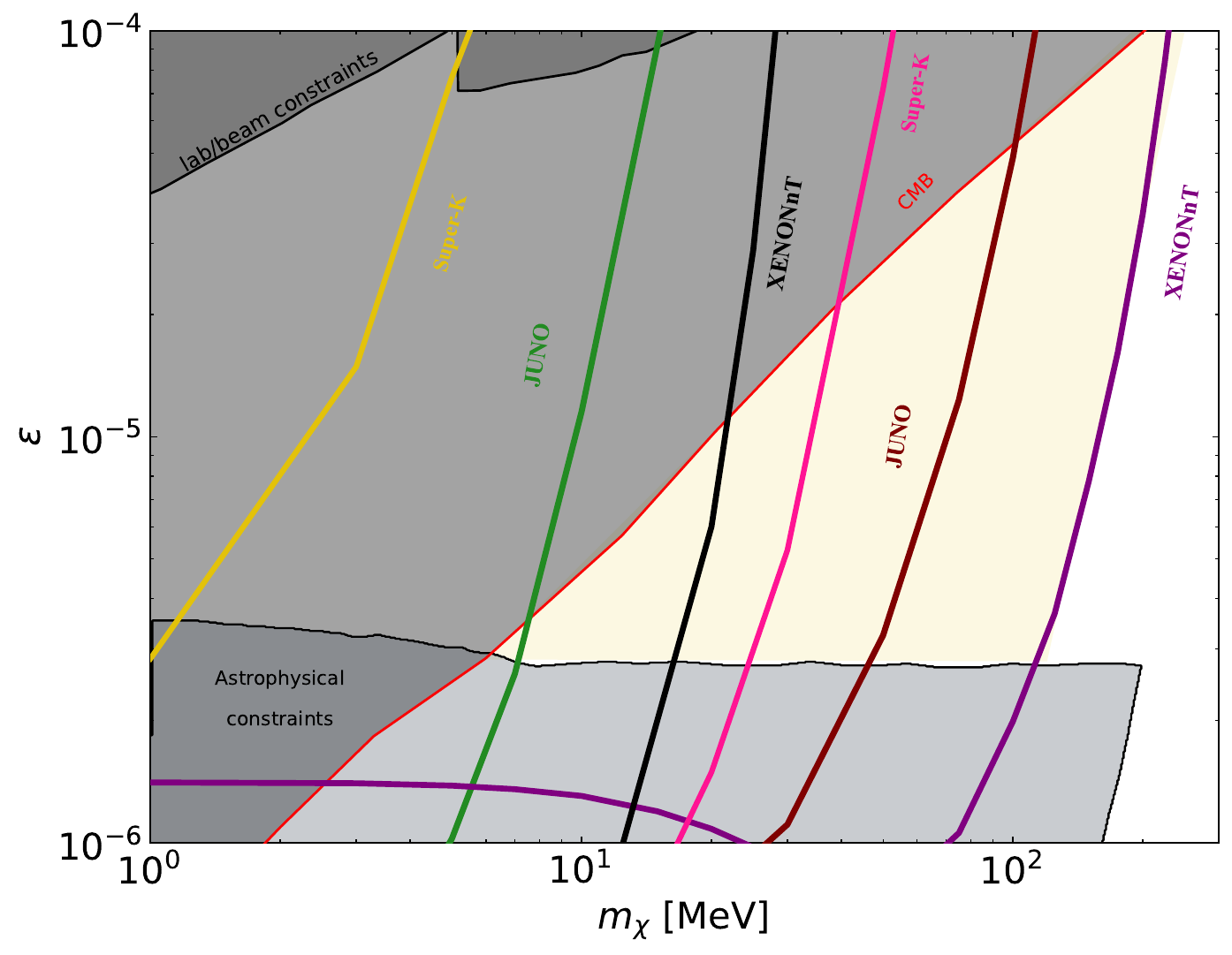}
\caption{\justifying
Exclusion region obtained from the electron recoil analysis.
     For PBHs with initial mass $ M_i=10^{15}~\rm g$, we obtained the contours
    of JUNO~(maroon), XENONnT~(purple), and
     Super-K~(magenta). For PBHs with initial mass
     $ M_i=8\times10^{15}~\rm g$, the contours are JUNO~(green),
     XENONnT~(black), and Super-K~(golden yellow). The yellow shaded region is the free parameter space this model excludes.
}
\label{fig:mcp_exclusion}
\end{figure}

The differential electron recoil rate is then,

\begin{equation}
\frac{dR}{dE_e}
=
\int dT_{\chi,f}\,
\frac{d\Phi_\chi^{\rm det}}{dT_{\chi,f}}\,
\frac{d\sigma}{dE_e},
\label{eq:electron_recoil_rate}
\end{equation}

where  $\frac{d\Phi_\chi^{\rm det}}{dT_{\chi,f}}$ is the detector flux. 

Assuming the shell structure of xenon, the differential cross section is given by~\cite{Harnik:2020ugb},

\begin{equation}
\frac{d\sigma_{\chi e}^{(b)}}{dE_e}
=
\left\{
\begin{aligned}
&
\frac{2\pi\alpha^2\epsilon^2}
{m_e\beta_\chi^2 E_b^2}
\left[
1-\beta_\chi^2\frac{E_b}{E_e^{\max}}
+\frac{E_b^2}{2E_\chi^2}
\right],
\\[1mm]
&\qquad E_b\leq E_e<2E_b,
\\[3mm]
&
\frac{2\pi\alpha^2\epsilon^2}
{m_e\beta_\chi^2(E_e-E_b)^2}
\left[
1-\beta_\chi^2\frac{E_e-E_b}{E_e^{\max}}
+\frac{(E_e-E_b)^2}{2E_\chi^2}
\right],
\\[1mm]
&\qquad E_e\geq2E_b,
\\[3mm]
&0,
\qquad E_e<E_b .
\end{aligned}
\right.
\label{eq:xe_shell_cross_section}
\end{equation}

Here, $\alpha$ is the fine structure constant, $E_b$ is the binding energy of the electron in shell $b$, $E_e$ is the total recoil energy deposited in xenon, and $E_e^{\rm max}$ is the maximum recoil energy the particle can transfer. The total differential scattering cross section after summing over all the shells $b$ is,

\begin{equation}
\frac{d\sigma_{\chi e}^{\rm Xe}}{dE_e}
=
\sum_b N_b\,
\frac{d\sigma_{\chi e}^{(b)}}{dE_e},
\end{equation}

where $N_b$ is the number of electrons occupied in $b$ shell.

For JUNO and Super-K, we use the free electron approximation for which the differential scattering cross section is,

\begin{equation}
\frac{d\sigma}{dE_e}
=
\frac{2\pi\alpha^2\epsilon^2}
{m_e\beta_\chi^2 E_e^2}
\left[
1
-\beta_\chi^2\frac{E_e}{E_e^{\max}}
+\frac{E_e^2}{2E_\chi^2}
\right].
\label{eq:free_electron_cross_section}
\end{equation}

The expected number of events in an electron-recoil energy bin is then,

\begin{equation}
N_{\rm bin}
=
\mathcal{E}\,n_{\rm target}
\int_{E_{\min}}^{E_{\max}}
dE_e\,
\eta(E_e)\frac{dR}{dE_e},
\label{eq:events_bin}
\end{equation}

where \(\mathcal{E}\) is the detector exposure, $n_{\rm target}$ is the number of scattering targets per unit detector mass, and $\eta(E_e)$ is the detection efficiency. In our calculations, we consider only downward-going MCPs.

We use the digitised signal efficiencies from Fig.~2 and the science data and total best-fit spectra from Fig.~5 of Ref.~\cite{XENON:2026rbz} For XENONnT, and perform a binned likelihood method for 1-140 keV recoil energy range. The SR0 and SR1b datasets have exposure of 1.09 t-yr and 1.37 t-yr, respectively. Comparing the observed science data with the total best-fit and with the expectation including the predicted MCP signal, we derive a \(90\%\) C.L. constraint.

For JUNO, we performed the electron recoil analysis with recoil energy bins 2-3, 3-5, and 5-16~MeV, with corresponding signal efficiencies $0.51$, $0.41$, and $0.52$ and exposure of $79$, $122$, and $162~\rm kt-yr$ respectively~\cite{JUNO:2020hqc}. Using a $\chi^2$ analysis, we compared the $\rm solar~neutrino+background$ events~\cite{JUNO:2020hqc} with the predicted events that include the expected MCP signal, and obtained a projected $90\%$ C.L. sensitivity.

We follow Ref.~\cite{Du:2022hms} for Super-K analysis, adopt their background-agnostic method and obtain the $90\%$ C.L. contour. For SK-I, SK-II, and SK-III, we use 16-88~MeV recoil energy range divided into 18 bins with an exposure of 176 kt-yr. The SK-IV contains 31 energy bins with 16-78 recoil energy range and 183 kt-yr exposure. The digitised data of the observed electron-like events and the efficiency are taken from~\cite{Super-Kamiokande:2011lwo} for SK-I, SK-II and SK-III, while for SK-IV we follow~\cite{Super-Kamiokande:2021jaq}.

\section{Result and conclusion}
\label{sec:result}

In Fig.~2, we show the exclusion limits obtained from the electron recoil analyses of XENONnT, JUNO, and Super-K for the two PBH populations. The Hawking emission of heavier MCPs becomes increasingly suppressed; hence, the flux decreases, and the lower boundary rises steeply.

For the PBHs with initial mass $8\times10^{15}~{\rm g}$, the bounds are restricted to relatively low MCP masses because these PBHs have a lower Hawking temperature which suppresses the emission of heavier MCPs strongly. Since XENONnT has a low recoil threshold, it can detect MCPs with relatively low kinetic energies, and extends to the highest MCP masses whereas JUNO and Super-K require increasingly energetic MCPs to produce observable recoil signals.

For the PBHs with initial mass $10^{15}~{\rm g}$, the constraints extend to significantly larger MCP masses. Since these PBHs have a higher Hawking temperature, the emission of heavier MCPs is less suppressed. XENONnT again extends to the largest MCP mass and imposes strong constraints in the relevant parameter space.

We neglect the supernova shock in our calculation. Although shocks may accelerate or evacuate part of the MCP population from the Galactic disk, PBHs continuously inject new MCPs throughout the galactic halo. Supernova shocks, therefore, cannot permanently deplete the entire population, and their effect is negligible over most of the parameter region considered.



\vspace{1em}

\begin{acknowledgments}
I thank Nirmal Raj and Ranjan Laha for an extensive and thorough discussion of the paper. I also acknowledge the assistance of OpenAI's GPT-6 model in code development.
\end{acknowledgments}


\bibliography{refs}

@article{Bertone:2004pz,
    author = "Bertone, Gianfranco and Hooper, Dan and Silk, Joseph",
    title = "{Particle dark matter: Evidence, candidates and constraints}",
    eprint = "hep-ph/0404175",
    archivePrefix = "arXiv",
    reportNumber = "FERMILAB-PUB-04-047-A",
    doi = "10.1016/j.physrep.2004.08.031",
    journal = "Phys. Rept.",
    volume = "405",
    pages = "279--390",
    year = "2005"
}

@article{Green:2020jor,
    author = "Green, Anne M. and Kavanagh, Bradley J.",
    title = "{Primordial Black Holes as a dark matter candidate}",
    eprint = "2007.10722",
    archivePrefix = "arXiv",
    primaryClass = "astro-ph.CO",
    doi = "10.1088/1361-6471/abc534",
    journal = "J. Phys. G",
    volume = "48",
    number = "4",
    pages = "043001",
    year = "2021"
}

@article{Tamta:2024pow,
    author = "Tamta, Manish and Raj, Nirmal and Sharma, Prateek",
    title = "{Entering the window of primordial black hole dark matter with x-ray microlensing}",
    eprint = "2405.20365",
    archivePrefix = "arXiv",
    primaryClass = "astro-ph.HE",
    doi = "10.1103/PhysRevD.111.043043",
    journal = "Phys. Rev. D",
    volume = "111",
    number = "4",
    pages = "043043",
    year = "2025"
}

@article{Saha:2025wgg,
    author = "Saha, Akash Kumar",
    title = "{Searching for dark matter with MeVCube}",
    eprint = "2501.13162",
    archivePrefix = "arXiv",
    primaryClass = "hep-ph",
    doi = "10.1103/15dd-mw83",
    journal = "Phys. Rev. D",
    volume = "111",
    number = "12",
    pages = "123041",
    year = "2025"
}

@article{Laha:2018zav,
    author = "Laha, Ranjan",
    title = "{Lensing of fast radio bursts: Future constraints on primordial black hole density with an extended mass function and a new probe of exotic compact fermion and boson stars}",
    eprint = "1812.11810",
    archivePrefix = "arXiv",
    primaryClass = "astro-ph.CO",
    doi = "10.1103/PhysRevD.102.023016",
    journal = "Phys. Rev. D",
    volume = "102",
    number = "2",
    pages = "023016",
    year = "2020"
}

@misc{koechler2026constrainingdarkmatterusing,
      title={Constraining dark matter using 20-year INTEGRAL/IBIS observations I: Primordial black holes}, 
      author={Jordan Koechler and Pedro De la Torre Luque},
      year={2026},
      eprint={2609.12044},
      archivePrefix={arXiv},
      primaryClass={astro-ph.HE},
      url={https://arxiv.org/abs/2609.12044}, 
}

@article{Saha:2024ies,
    author = "Saha, Akash Kumar and Singh, Abhijeet and Parashari, Priyank and Laha, Ranjan",
    title = "{Hunting primordial black hole dark matter in the Lyman-$\alpha $ forest}",
    eprint = "2409.10617",
    archivePrefix = "arXiv",
    primaryClass = "astro-ph.CO",
    doi = "10.1140/epjc/s10052-025-14827-1",
    journal = "Eur. Phys. J. C",
    volume = "85",
    number = "10",
    pages = "1117",
    year = "2025"
}

@article{Saha:2021pqf,
    author = "Saha, Akash Kumar and Laha, Ranjan",
    title = "{Sensitivities on nonspinning and spinning primordial black hole dark matter with global 21-cm troughs}",
    eprint = "2112.10794",
    archivePrefix = "arXiv",
    primaryClass = "astro-ph.CO",
    doi = "10.1103/PhysRevD.105.103026",
    journal = "Phys. Rev. D",
    volume = "105",
    number = "10",
    pages = "103026",
    year = "2022"
}

@article{Laha:2019ssq,
    author = "Laha, Ranjan",
    title = "{Primordial Black Holes as a Dark Matter Candidate Are Severely Constrained by the Galactic Center 511 keV $\gamma$ -Ray Line}",
    eprint = "1906.09994",
    archivePrefix = "arXiv",
    primaryClass = "astro-ph.HE",
    reportNumber = "CERN-TH-2019-099",
    doi = "10.1103/PhysRevLett.123.251101",
    journal = "Phys. Rev. Lett.",
    volume = "123",
    number = "25",
    pages = "251101",
    year = "2019"
}

@article{Dasgupta:2019cae,
    author = "Dasgupta, Basudeb and Laha, Ranjan and Ray, Anupam",
    title = "{Neutrino and positron constraints on spinning primordial black hole dark matter}",
    eprint = "1912.01014",
    archivePrefix = "arXiv",
    primaryClass = "hep-ph",
    reportNumber = "CERN-TH-2019-212, TIFR/TH/19-40",
    doi = "10.1103/PhysRevLett.125.101101",
    journal = "Phys. Rev. Lett.",
    volume = "125",
    number = "10",
    pages = "101101",
    year = "2020"
}

@article{Laha:2020ivk,
    author = "Laha, Ranjan and Mu{\~n}oz, Julian B. and Slatyer, Tracy R.",
    title = "{INTEGRAL constraints on primordial black holes and particle dark matter}",
    eprint = "2004.00627",
    archivePrefix = "arXiv",
    primaryClass = "astro-ph.CO",
    doi = "10.1103/PhysRevD.101.123514",
    journal = "Phys. Rev. D",
    volume = "101",
    number = "12",
    pages = "123514",
    year = "2020"
}

@article{Laha:2020vhg,
    author = "Laha, Ranjan and Lu, Philip and Takhistov, Volodymyr",
    title = "{Gas heating from spinning and non-spinning evaporating primordial black holes}",
    eprint = "2009.11837",
    archivePrefix = "arXiv",
    primaryClass = "astro-ph.CO",
    reportNumber = "CERN-TH-2020-159, IPMU21-0004",
    doi = "10.1016/j.physletb.2021.136459",
    journal = "Phys. Lett. B",
    volume = "820",
    pages = "136459",
    year = "2021"
}

@article{Ray:2021mxu,
    author = "Ray, Anupam and Laha, Ranjan and Mu{\~n}oz, Julian B. and Caputo, Regina",
    title = "{Near future MeV telescopes can discover asteroid-mass primordial black hole dark matter}",
    eprint = "2102.06714",
    archivePrefix = "arXiv",
    primaryClass = "astro-ph.CO",
    reportNumber = "TIFR/TH/21-1",
    doi = "10.1103/PhysRevD.104.023516",
    journal = "Phys. Rev. D",
    volume = "104",
    number = "2",
    pages = "023516",
    year = "2021"
}

@article{Fick:1855,
    author  = {Fick, Adolph},
    title   = {V. On liquid diffusion},
    journal = {The London, Edinburgh, and Dublin Philosophical Magazine and Journal of Science},
    volume  = {10},
    number  = {63},
    pages   = {30--39},
    year    = {1855},
    doi     = {10.1080/14786445508641925}
}

@article{10.1093/mnras/152.1.75,
    author = {Hawking, Stephen},
    title = {Gravitationally Collapsed Objects of Very Low Mass},
    journal = {Monthly Notices of the Royal Astronomical Society},
    volume = {152},
    number = {1},
    pages = {75-78},
    year = {1971},
    month = {04},
    issn = {0035-8711},
    doi = {10.1093/mnras/152.1.75},
    url = {https://doi.org/10.1093/mnras/152.1.75},
    eprint = {https://academic.oup.com/mnras/article-pdf/152/1/75/9360899/mnras152-0075.pdf},
}

@article{Cirelli:2024ssz,
    author = "Cirelli, Marco and Strumia, Alessandro and Zupan, Jure",
    title = "{Dark Matter}",
    eprint = "2406.01705",
    archivePrefix = "arXiv",
    primaryClass = "hep-ph",
    doi = "10.21468/SciPostPhysRev.1",
    journal = "SciPost Phys. Rev.",
    volume = "1",
    year = "2026"
}

@article{Prinz_1998,
   title={Search for Millicharged Particles at SLAC},
   volume={81},
   ISSN={1079-7114},
   url={http://dx.doi.org/10.1103/PhysRevLett.81.1175},
   DOI={10.1103/physrevlett.81.1175},
   number={6},
   journal={Physical Review Letters},
   publisher={American Physical Society (APS)},
   author={Prinz, A. A. and Baggs, R. and Ballam, J. and Ecklund, S. and Fertig, C. and Jaros, J. A. and Kase, K. and Kulikov, A. and Langeveld, W. G. J. and Leonard, R. and Marvin, T. and Nakashima, T. and Nelson, W. R. and Odian, A. and Pertsova, M. and Putallaz, G. and Weinstein, A.},
   year={1998},
   month=Aug, pages={1175–1178} }

@article{DelNobile:2015bqo,
    author = "Del Nobile, Eugenio and Nardecchia, Marco and Panci, Paolo",
    title = "{Millicharge or Decay: A Critical Take on Minimal Dark Matter}",
    eprint = "1512.05353",
    archivePrefix = "arXiv",
    primaryClass = "hep-ph",
    reportNumber = "DFPD-2015-TH-29",
    doi = "10.1088/1475-7516/2016/04/048",
    journal = "JCAP",
    volume = "04",
    pages = "048",
    year = "2016"
}

@article{Liu:2019knx,
    author = "Liu, Hongwan and Outmezguine, Nadav Joseph and Redigolo, Diego and Volansky, Tomer",
    title = "{Reviving Millicharged Dark Matter for 21-cm Cosmology}",
    eprint = "1908.06986",
    archivePrefix = "arXiv",
    primaryClass = "hep-ph",
    reportNumber = "MIT-CTP/5126",
    doi = "10.1103/PhysRevD.100.123011",
    journal = "Phys. Rev. D",
    volume = "100",
    number = "12",
    pages = "123011",
    year = "2019"
}

@article{Krumholz:2022oja,
    author = "Krumholz, Mark R. and Crocker, Roland M. and Offner, Stella S. R.",
    title = "{The cosmic ray ionization and {\ensuremath{\gamma}}-ray budgets of star-forming galaxies}",
    eprint = "2211.03488",
    archivePrefix = "arXiv",
    primaryClass = "astro-ph.GA",
    doi = "10.1093/mnras/stad459",
    journal = "Mon. Not. Roy. Astron. Soc.",
    volume = "520",
    number = "4",
    pages = "5126--5143",
    year = "2023"
}

@article{Pospelov:2020ktu,
    author = "Pospelov, Maxim and Ramani, Harikrishnan",
    title = "{Earth-bound millicharge relics}",
    eprint = "2012.03957",
    archivePrefix = "arXiv",
    primaryClass = "hep-ph",
    doi = "10.1103/PhysRevD.103.115031",
    journal = "Phys. Rev. D",
    volume = "103",
    number = "11",
    pages = "115031",
    year = "2021"
}

@article{PhysRevD.41.2388,
  title = {Getting a charge out of dark matter},
  author = {Dimopoulos, Savas and Eichler, David and Esmailzadeh, Rahim and Starkman, Glenn D.},
  journal = {Phys. Rev. D},
  volume = {41},
  issue = {8},
  pages = {2388--2397},
  numpages = {0},
  year = {1990},
  month = {Apr},
  publisher = {American Physical Society},
  doi = {10.1103/PhysRevD.41.2388},
  url = {https://link.aps.org/doi/10.1103/PhysRevD.41.2388}
}

@article{McDermott:2010pa,
    author = "McDermott, Samuel D. and Yu, Hai-Bo and Zurek, Kathryn M.",
    title = "{Turning off the Lights: How Dark is Dark Matter?}",
    eprint = "1011.2907",
    archivePrefix = "arXiv",
    primaryClass = "hep-ph",
    reportNumber = "MCTP-10-52",
    doi = "10.1103/PhysRevD.83.063509",
    journal = "Phys. Rev. D",
    volume = "83",
    pages = "063509",
    year = "2011"
}

@article{Carr:2020gox,
    author = "Carr, Bernard and Kohri, Kazunori and Sendouda, Yuuiti and Yokoyama, Jun'ichi",
    title = "{Constraints on primordial black holes}",
    eprint = "2002.12778",
    archivePrefix = "arXiv",
    primaryClass = "astro-ph.CO",
    reportNumber = "RESCEU-03/20; KEK-Cosmo-249; KEK-TH-2199; IPMU20-0024",
    doi = "10.1088/1361-6633/ac1e31",
    journal = "Rept. Prog. Phys.",
    volume = "84",
    number = "11",
    pages = "116902",
    year = "2021"
}

@article{Magill:2018tbb,
    author = "Magill, Gabriel and Plestid, Ryan and Pospelov, Maxim and Tsai, Yu-Dai",
    title = "{Millicharged particles in neutrino experiments}",
    eprint = "1806.03310",
    archivePrefix = "arXiv",
    primaryClass = "hep-ph",
    reportNumber = "FERMILAB-PUB-18-631-A",
    doi = "10.1103/PhysRevLett.122.071801",
    journal = "Phys. Rev. Lett.",
    volume = "122",
    number = "7",
    pages = "071801",
    year = "2019"
}

@article{Chuzhoy:2008zy,
    author = "Chuzhoy, Leonid and Kolb, Edward W.",
    title = "{Reopening the window on charged dark matter}",
    eprint = "0809.0436",
    archivePrefix = "arXiv",
    primaryClass = "astro-ph",
    doi = "10.1088/1475-7516/2009/07/014",
    journal = "JCAP",
    volume = "07",
    pages = "014",
    year = "2009"
}

@article{Calabrese:2021src,
    author = "Calabrese, Roberta and Chianese, Marco and Fiorillo, Damiano F. G. and Saviano, Ninetta",
    title = "{Direct detection of light dark matter from evaporating primordial black holes}",
    eprint = "2107.13001",
    archivePrefix = "arXiv",
    primaryClass = "hep-ph",
    doi = "10.1103/PhysRevD.105.L021302",
    journal = "Phys. Rev. D",
    volume = "105",
    number = "2",
    pages = "L021302",
    year = "2022"
}

@article{Chang:2018rso,
    author = "Chang, Jae Hyeok and Essig, Rouven and McDermott, Samuel D.",
    title = "{Supernova 1987A Constraints on Sub-GeV Dark Sectors, Millicharged Particles, the QCD Axion, and an Axion-like Particle}",
    eprint = "1803.00993",
    archivePrefix = "arXiv",
    primaryClass = "hep-ph",
    reportNumber = "YITP-SB-18-01, FERMILAB-PUB-17-432-T",
    doi = "10.1007/JHEP09(2018)051",
    journal = "JHEP",
    volume = "09",
    pages = "051",
    year = "2018"
}

@article{ParticleDataGroup:2026mpi,
    author = "Takahashi, F. and others",
    collaboration = "Particle Data Group",
    title = "{Review of Particle Physics}",
    doi = "10.1142/s0217751x26300115",
    journal = "Int. J. Mod. Phys. A",
    volume = "41",
    number = "22",
    pages = "2630011",
    year = "2026"
}

@article{Li:2024pcp,
    author = "Li, Changqian and Liu, Zuowei and Lu, Wenxi and Ye, Zicheng",
    title = "{Low-energy supernova constraints on millicharged particles}",
    eprint = "2408.04953",
    archivePrefix = "arXiv",
    primaryClass = "hep-ph",
    doi = "10.1007/JHEP07(2025)116",
    journal = "JHEP",
    volume = "07",
    pages = "116",
    year = "2025"
}

@article{Vinyoles:2015khy,
    author = "Vinyoles, N{\'u}ria and Vogel, Hendrik",
    title = "{Minicharged Particles from the Sun: A Cutting-Edge Bound}",
    eprint = "1511.01122",
    archivePrefix = "arXiv",
    primaryClass = "hep-ph",
    reportNumber = "MPP-2015-199",
    doi = "10.1088/1475-7516/2016/03/002",
    journal = "JCAP",
    volume = "03",
    pages = "002",
    year = "2016"
}

@article{Arbey:2019mbc,
    author = "Arbey, Alexandre and Auffinger, J{\'e}r{\'e}my",
    title = "{BlackHawk: A public code for calculating the Hawking evaporation spectra of any black hole distribution}",
    eprint = "1905.04268",
    archivePrefix = "arXiv",
    primaryClass = "gr-qc",
    reportNumber = "CERN-TH-2019-067",
    doi = "10.1140/epjc/s10052-019-7161-1",
    journal = "Eur. Phys. J. C",
    volume = "79",
    number = "8",
    pages = "693",
    year = "2019"
}

@article{Dunsky:2018mqs,
    author = "Dunsky, David and Hall, Lawrence J. and Harigaya, Keisuke",
    title = "{CHAMP Cosmic Rays}",
    eprint = "1812.11116",
    archivePrefix = "arXiv",
    primaryClass = "astro-ph.HE",
    doi = "10.1088/1475-7516/2019/07/015",
    journal = "JCAP",
    volume = "07",
    pages = "015",
    year = "2019"
}

@article{Harnik:2020ugb,
    author = "Harnik, Roni and Plestid, Ryan and Pospelov, Maxim and Ramani, Harikrishnan",
    title = "{Millicharged cosmic rays and low recoil detectors}",
    eprint = "2010.11190",
    archivePrefix = "arXiv",
    primaryClass = "hep-ph",
    reportNumber = "FERMILAB-PUB-20-523-T",
    doi = "10.1103/PhysRevD.103.075029",
    journal = "Phys. Rev. D",
    volume = "103",
    number = "7",
    pages = "075029",
    year = "2021"
}

@article{Kachelriess:2020ams,
    author = "Kachelrie{\ss}, M. and Tjemsland, J.",
    title = "{Reacceleration of charged dark matter}",
    eprint = "2006.10479",
    archivePrefix = "arXiv",
    primaryClass = "astro-ph.HE",
    doi = "10.1088/1475-7516/2020/10/001",
    journal = "JCAP",
    volume = "10",
    pages = "001",
    year = "2020"
}

@article{Strong_2007,
   title={Cosmic-Ray Propagation and Interactions in the Galaxy},
   volume={57},
   ISSN={1545-4134},
   url={http://dx.doi.org/10.1146/annurev.nucl.57.090506.123011},
   DOI={10.1146/annurev.nucl.57.090506.123011},
   number={1},
   journal={Annual Review of Nuclear and Particle Science},
   publisher={Annual Reviews},
   author={Strong, Andrew W. and Moskalenko, Igor V. and Ptuskin, Vladimir S.},
   year={2007},
   month=Nov, pages={285–327} }

@misc{Perri:2025tvq,
    author = "Perri, Daniele and Farrar, Glennys",
    title = "{The window on heavy charged dark matter was never open}",
    eprint = "2510.17026",
    archivePrefix = "arXiv",
    primaryClass = "hep-ph",
    month = "10",
    year = "2025"
}

@article{Pakmor:2020,
    author = "Pakmor, Ruediger and van de Voort, Freeke and Bieri, Rebekka
              and Gomez, Facundo A. and Grand, Robert J. J. and Guillet, Thomas
              and Marinacci, Federico and Pfrommer, Christoph
              and Simpson, Christine M. and Springel, Volker",
    title = "{Magnetising the circumgalactic medium of disk galaxies}",
    eprint = "1911.11163",
    archivePrefix = "arXiv",
    primaryClass = "astro-ph.GA",
    doi = "10.1093/mnras/staa2530",
    journal = "Mon. Not. Roy. Astron. Soc.",
    volume = "498",
    number = "3",
    pages = "3125--3137",
    year = "2020"
}

@article{AlvesBatista:2022,
    author = "Alves Batista, Rafael and Becker Tjus, Julia and Dörner, Julien
              and Dundovic, Andrej and Eichmann, Björn and Frie, Antonius
              and Heiter, Christopher and Hoerbe, Mario R. and Kampert, Karl-Heinz
              and Merten, Lukas and Müller, Gero and Reichherzer, Patrick
              and Saveliev, Andrey and Schlegel, Leander and Sigl, Günter
              and van Vliet, Arjen and Winchen, Tobias",
    title = "{CRPropa 3.2 -- an advanced framework for high-energy particle
              propagation in extragalactic and galactic spaces}",
    eprint = "2208.00107",
    archivePrefix = "arXiv",
    primaryClass = "astro-ph.HE",
    doi = "10.1088/1475-7516/2022/09/035",
    journal = "JCAP",
    volume = "09",
    pages = "035",
    year = "2022"
}

@article{Merten:2017,
    author = "Merten, Lukas and Becker Tjus, Julia and Fichtner, Horst
              and Eichmann, Björn and Sigl, Günter",
    title = "{CRPropa 3.1 -- A low energy extension based on stochastic
              differential equations}",
    eprint = "1704.07484",
    archivePrefix = "arXiv",
    primaryClass = "astro-ph.IM",
    doi = "10.1088/1475-7516/2017/06/046",
    journal = "JCAP",
    volume = "06",
    pages = "046",
    year = "2017"
}

@article{JUNO:2020hqc,
    author = "Abusleme, Angel and others",
    collaboration = "JUNO",
    title = "{Feasibility and physics potential of detecting $^8$B solar neutrinos at JUNO}",
    eprint = "2006.11760",
    archivePrefix = "arXiv",
    primaryClass = "hep-ex",
    doi = "10.1088/1674-1137/abd92a",
    journal = "Chin. Phys. C",
    volume = "45",
    number = "2",
    pages = "023004",
    year = "2021"
}

@article{Du:2022hms,
    author = "Du, Mingxuan and Fang, Rundong and Liu, Zuowei",
    title = "{Millicharged particles from proton bremsstrahlung in the atmosphere}",
    eprint = "2211.11469",
    archivePrefix = "arXiv",
    primaryClass = "hep-ph",
    doi = "10.1007/JHEP08(2024)174",
    journal = "JHEP",
    volume = "08",
    pages = "174",
    year = "2024"
}

@article{Super-Kamiokande:2011lwo,
    author = "Bays, K. and others",
    collaboration = "Super-Kamiokande",
    title = "{Supernova Relic Neutrino Search at Super-Kamiokande}",
    eprint = "1111.5031",
    archivePrefix = "arXiv",
    primaryClass = "hep-ex",
    doi = "10.1103/PhysRevD.85.052007",
    journal = "Phys. Rev. D",
    volume = "85",
    pages = "052007",
    year = "2012"
}

@article{Super-Kamiokande:2021jaq,
    author = "Abe, K. and others",
    collaboration = "Super-Kamiokande",
    title = "{Diffuse supernova neutrino background search at Super-Kamiokande}",
    eprint = "2109.11174",
    archivePrefix = "arXiv",
    primaryClass = "astro-ph.HE",
    doi = "10.1103/PhysRevD.104.122002",
    journal = "Phys. Rev. D",
    volume = "104",
    number = "12",
    pages = "122002",
    year = "2021"
}

@misc{XENON:2026rbz,
    author = "Aprile, E. and others",
    collaboration = "XENON",
    title = "{First Measurement of Solar Neutrinos through Elastic Neutrino-Electron Scattering at the keV Scale}",
    eprint = "2608.29450",
    archivePrefix = "arXiv",
    primaryClass = "hep-ex",
    month = "8",
    year = "2026"
}

@misc{Berger:PSTAR,
    author = {Berger, M. J. and Coursey, J. S. and Zucker, M. A. and Chang, J.},
    title  = {{ESTAR, PSTAR, and ASTAR: Computer Programs for Calculating Stopping-Power and Range Tables for Electrons, Protons, and Helium Ions}},
    year = {2005},
    note  = {Version 1.2.3, NIST Standard Reference Database 124,
    National Institute of Standards and Technology},
    doi  = {10.18434/T4NC7P},
    url = {https://physics.nist.gov/Star},
    urldate= {2026-08-20}
    }

\end{document}